\documentstyle[psfig,twocolumn,aps]{revtex}
\begin{document}
\draft

%
\twocolumn[\hsize\textwidth\columnwidth\hsize\csname@twocolumnfalse\endcsname
%
%

\title{From the Hubbard to the $SO(5)$ Ladder: A Numerical Study}

\author{Daniel Duffy$^1$, Stephan Haas$^2$ and Eugene Kim$^1$}

\address{$^1$Department of Physics, University of California, 
Santa Barbara CA 93106;\\
$^2$Theoretische Physik, ETH-H\"onggerberg, CH-8093 Zurich, Switzerland
}

\date{\today}
\maketitle

\begin{abstract}
The Hubbard Hamiltonian on a two-leg ladder is studied numerically
using quantum Monte Carlo and Exact Diagonalization techniques.  A
rung interaction, $V$, is turned on such that the resulting model has
an exact $SO(5)$ symmetry when $V=-U$.  The evolution of the low
energy excitation spectrum is presented from the pure Hubbard ladder
to the $SO(5)$ ladder.  It is shown that the low energy excitations in
the pure Hubbard ladder have an approximate $SO(5)$ symmetry.
\end{abstract}

\pacs{PACS numbers: 71.10.-w,71.10.Fd,74.20.-z}
\vskip2pc]
\narrowtext

In recent years, various compounds containing two-leg cuprate ladders have
been synthesized\cite{elbiorice}. The ground states of these materials
are characterized by strong electronic interactions with a variety of
competing ground states.  Theoretically, the focus has been on using
the Hubbard\cite{noack} and $t-J$ models\cite{troyer} to understand
the ground state properties of these materials.  At half-filling, the
two-leg ladders exhibit a spin-gap insulating phase, while away from
half-filling $d_{x^2-y^2}$ pairing and charge density wave (CDW)
correlations become enhanced\cite{hayward}.

Recently, Zhang proposed a model with $SO(5)$ symmetry to relate the
two phases of antiferromagnetism (AF) and $d$-wave superconductivity
(dSC), even though these two phases have quite distinct ground
states\cite{zhang}. Subsequently, several groups have constructed
microscopic models which have an exact $SO(5)$
symmetry\cite{szh,rabello,henley,burgess}. Furthermore, it has been
argued that both the Hubbard\cite{meixner} and $t-J$ models\cite{eder}
in two-dimensions (2D) have approximate $SO(5)$ symmetry.

In this paper, we examine one such microscopic Hamiltonian\cite{szh}
using quantum Monte Carlo (QMC) and exact diagonalization (ED)
techniques.  The model we study can be changed adiabatically from the
pure Hubbard model on a two-leg ladder to a model with $SO(5)$
symmetry, thus elucidating how the low energy spectrum evolves as the
higher symmetry is approached.  In this way, we can test whether the
low lying excitations of the Hubbard ladder are well described by the
$SO(5)$ picture.

The microscopic $SO(5)$ ladder we investigate here was first introduced by
Scalapino, Zhang and Hanke (SZH)\cite{szh}. The SZH model contains
only local interactions on a rung of the ladder and is therefore much
easier to visualize and implement numerically than 2D $SO(5)$
models with long range interactions.  The Hamiltonian is given by

\begin{eqnarray}
H = H_{hop} + H_{int}.
\end{eqnarray}

\noindent $H_{hop}$ allows electrons to move along the
ladder (in the $\hat{x}$ direction) as well as within a rung (in the 
$\hat{y}$ direction):

\begin{eqnarray}
H_{hop} = -t_{\parallel}{ \sum_{{\bf i},s}
       (c^{\dagger}_{{\bf i},s}c_{{\bf i+\hat{x}},s}+H.c.)} \nonumber\\
           -t_{\perp}{ \sum_{{\bf i},s}
       (c^{\dagger}_{{\bf i},s}c_{{\bf i+\hat{y}},s}+H.c.)}
\end{eqnarray}

\noindent where $c^\dagger_{{\bf i},s}$ creates an electron at site
${\bf i}$ with spin projection $s$.  The interaction term,
$H_{int}$, is given by

\begin{eqnarray}
H_{int} = U{ \sum_{{\bf{i}}}(n_{{\bf{i}} \uparrow}-1/2)( n_{{\bf{i}}
\downarrow}-1/2)+\mu\sum_{{\bf{i}},s}n_{{\bf{i}}s} } \nonumber \\
+ V \sum_{\bf i} (n_{\bf i}-1)(n_{\bf i+\hat{y}}-1)
+ J \sum_{\bf i} \vec{S}_{\bf i}\cdot\vec{S}_{\bf i+\hat{y}}
\end{eqnarray}

\noindent where $n_{\bf i}$ is the number operator for site ${\bf i}$;
$U$ is the Coulomb repulsion between electrons occupying the same
site; $V$ is an interaction between electrons on a given rung; $J$ is a
rung spin-spin interaction with

\begin{eqnarray}
\vec{S_{\bf i}} = \frac{1}{2}\sum_{s,s'} c^\dagger_{{\bf i},s} 
\vec{\sigma}_{s,s'}c_{{\bf i},s'}.
\end{eqnarray}

\noindent In order for this Hamiltonian to be manifestly $SO(5)$
symmetric, the condition of $J=4(U+V)$ must be imposed.

The phase diagram of the SZH model at half-filling has been obtained
at strong coupling, i.e., $U,V>>t_\perp,t_\parallel$\cite{szh}.
Predominately for $U>0$, the ground state is well described as a
product of rung singlets, i.e., a spin-gap insulator.  Doped holes in
this ground state form $d_{x^2-y^2}$ rung pairs and exhibit power law
pairing and CDW correlations.  A phase transition occurs for positive
values of $U$ from the singlet ground state into a triplet ground
state along the line $V=-U$, where each rung is occupied by a triplet
magnon or a doublet pair.  Furthermore, the triplet ground state can
be considered to be an $SO(5)$ generalization of the spin-one
Heisenberg chain, i.e., a system with a finite excitation gap and
short range correlations.

At weak couplings ($U,V << t_\perp,t_\parallel$), the phase diagram
of the SZH model was obtained using a perturbative renormalization
group analysis\cite{lin}. The resulting phase diagram, although
consistent with the strong coupling result, contains new phases not
found in the strong coupling limit.  Furthermore, it was shown
that two-leg Hubbard-like ladders flow to $SO(5)$ symmetry even when
explicit symmetry breaking terms were included, such as longer range
hoppings\cite{arrigoni}.

In this paper, we explicitly turn off the spin-spin interaction
by requiring that $J=0$\cite{qmcnote}. Thus, our model is $SO(5)$
symmetric {\em only} when $V=-U$, where the ground state at large
couplings was found to have a degeneracy between the rung singlet and
triplet $SO(5)$ multiplets.  Consequently, we now have an adjustable
model which interpolates from the Hubbard ladder to the $SO(5)$ ladder
by varying $V$ from 0 to $-U$.  We will study this model in order to
investigate how the low lying excitations evolve as $V\rightarrow -U$,
i.e., as the model approaches $SO(5)$ symmetry.  Throughout, we will
take the isotropic hopping case of $t = t_\perp = t_\parallel = 1$ and
will only consider $U>0$.

In Zhang's $SO(5)$ theory, the superspin vector takes on a fixed
magnitude below some characteristic temperature, $T^*$, and its
direction fluctuates between the AF and dSC phases.  For the case in
which the model is manifestly $SO(5)$ symmetric $(V=-U)$, $T^*$ is
pushed to infinity and all of the components of the superspin vector
will be equal.  To test this, we use deterministic quantum Monte Carlo
(QMC)\cite{qmc} to measure the first and fourth components of the
superspin vector given by

\begin{eqnarray}
n_1(r) &=& \frac{(-1)^r}{2}(\Delta^\dagger + \Delta) \nonumber \\
       &=& \frac{(-1)^r}{2}\bigl(-ic^\dagger(r) \sigma_y \
                             c^\dagger(r+\hat{y}) + h.c.\bigr)
\end{eqnarray}

\noindent and 

\begin{eqnarray}
n_4(r) = \frac{(-1)^r}{2}\bigl(c^\dagger(r) \sigma_z c(r)
         -c^\dagger(r+\hat{y}) \sigma_z c(r+\hat{y})\bigr).
\end{eqnarray}

\noindent Here $r$ is the rung index, and the spinor index on the
$c$ operators has been suppressed, i.e., $c(r) =
(c_{\uparrow,r},c_{\downarrow,r})$.

\begin{figure}[tbh]
\vspace{0.35cm}
\centerline{\psfig{figure=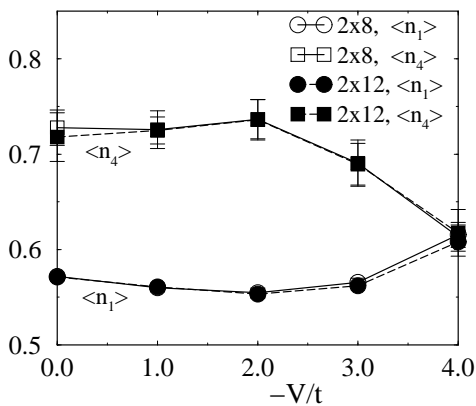,width=7cm,bbllx=98pt,bblly=36pt,bburx=590pt,bbury=700pt,angle=270}}
\caption{$\langle n_1\rangle$ (circles) and $\langle n_4\rangle$ (squares)
for $U=4t$ and $\beta t = 2$ at half-filling for the $2 \times 8$ 
(open symbols) and the $2 \times 12$ (filled symbols) lattice.}
\label{fig1}
\end{figure}

In Fig. 1, the sum of $n_1$ and $n_4$ over all rungs of a $2\times 8$
and $2\times 12$ ladder are shown, i.e., $\langle n_{1(4)}\rangle =
\frac{1}{N_{rungs}}\sum_r \langle n^\dagger_{1(4)}(r) n_{1(4)}(0)
\rangle$.  At half-filling $(\mu=0)$ for an intermediate strength
coupling of $U=4t$, the expected behavior is seen: as $-V$ approaches
the $SO(5)$ value of $U$, the correlations become equal.  Note that
even at half-filling a sign problem exists due to the new
Hubbard-Stratonovich fields introduced by the $V$ term\cite{qmcV}.  At
small to intermediate coupling strengths, i.e., $U<6t$, reliable
results can be obtained, even at the $SO(5)$ point when $V=-U$.
However, when $U$ becomes large, and consequently $-V$ becomes large
as the symmetric point is approached, the sign problem becomes
unmanageable.  Thus, only $U=4t$ with $\beta t = 2$ QMC results are
shown\cite{temp}.

Recall that at strong couplings along the phase transition line of
$U=-V$, the ground state was found to have a degeneracy between the
rung singlet and rung triplet $SO(5)$ multiplets.  Therefore, the
correlation functions should decay as a power law in distance, $r$,
with some thermal exponential activation, i.e., $n_{1(4)}(r,\beta)\sim
\mbox{exp}(-\beta/\xi)/r^\alpha$ (where $\xi$ is the thermal
correlation length).  However, when $V$ is different from $-U$, the
correlations should decay exponentially in $r$.

\begin{figure}[tbh]
\vspace{0.45cm}
\centerline{\psfig{figure=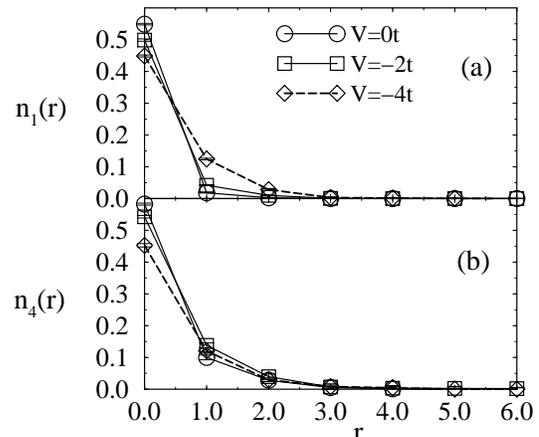,width=7cm,bbllx=98pt,bblly=36pt,bburx=590pt,bbury=700pt,angle=270}}
\caption{(a) $n_1(r)=\langle n_1^\dagger(r)n_1(0) \rangle$ for $U=4t$
and $\beta t = 2$ at half-filling on a $2\times 12$ ladder as a
function of distance along the ladder for three different values of
the rung attraction $V$.  Note that the results for the $SO(5)$ ladder
are for $V=-4t$ (open diamonds). (b) Same as in (a) for
$n_4(r)=\langle n_4^\dagger(r)n_4(0) \rangle$.}
\label{fig2}
\end{figure}

In order to test if this critical behavior could be seen, the correlations
of the superspin components were measured as a function of distance along
the ladder.  In Fig. 2, we show the results obtained for the $2\times 12$
ladder with $U=4t$ at half-filling as $V$ approaches the $SO(5)$ value.
In Fig. 2(a), a qualitative change in the correlations of $n_1(r)$
is clearly observed
between the Hubbard ladder and the $SO(5)$ ladder, indicating
a crossover from a power law to an exponential decay in $n_1(r)$ as the
$SO(5)$ point is approached.  In contrast,
the change in the $n_4(r)$ correlations (Fig. 2(b)) is less pronounced.

It was not possible to cleanly extract the power law exponents from a log-log
analysis of these correlations for several reasons.  First, $U=4t$ is 
not strong enough to recover the exact degeneracy between $SO(5)$
multiplets which is present only at large coupling strengths.  Also, at
a relatively high temperature of $\beta t = 2$, the thermal correlation
length tends to dominate the behavior of the superspin
components\cite{noack}.  Since reliable
measurements at either larger couplings or lower temperatures
could not be obtained, let us now turn to a zero-temperature
exact diagonalization (ED) analysis to better understand
the low energy spectrum\cite{ed}.

\begin{figure}[tbh]
\vspace{0.25cm}
\centerline{\psfig{figure=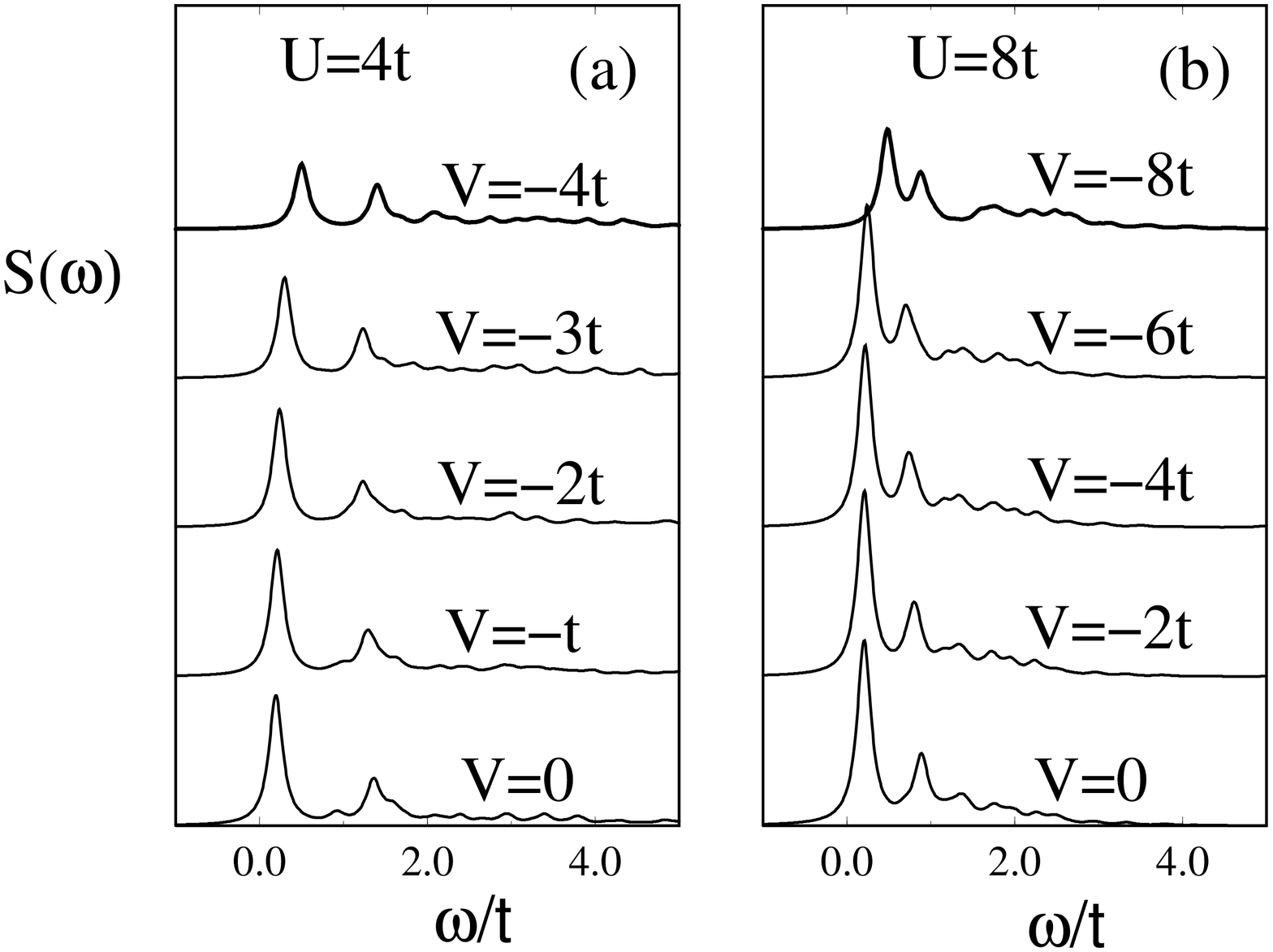,width=7cm,bbllx=98pt,bblly=36pt,bburx=590pt,bbury=700pt,angle=270}}
\vspace{0.5cm}
\centerline{\psfig{figure=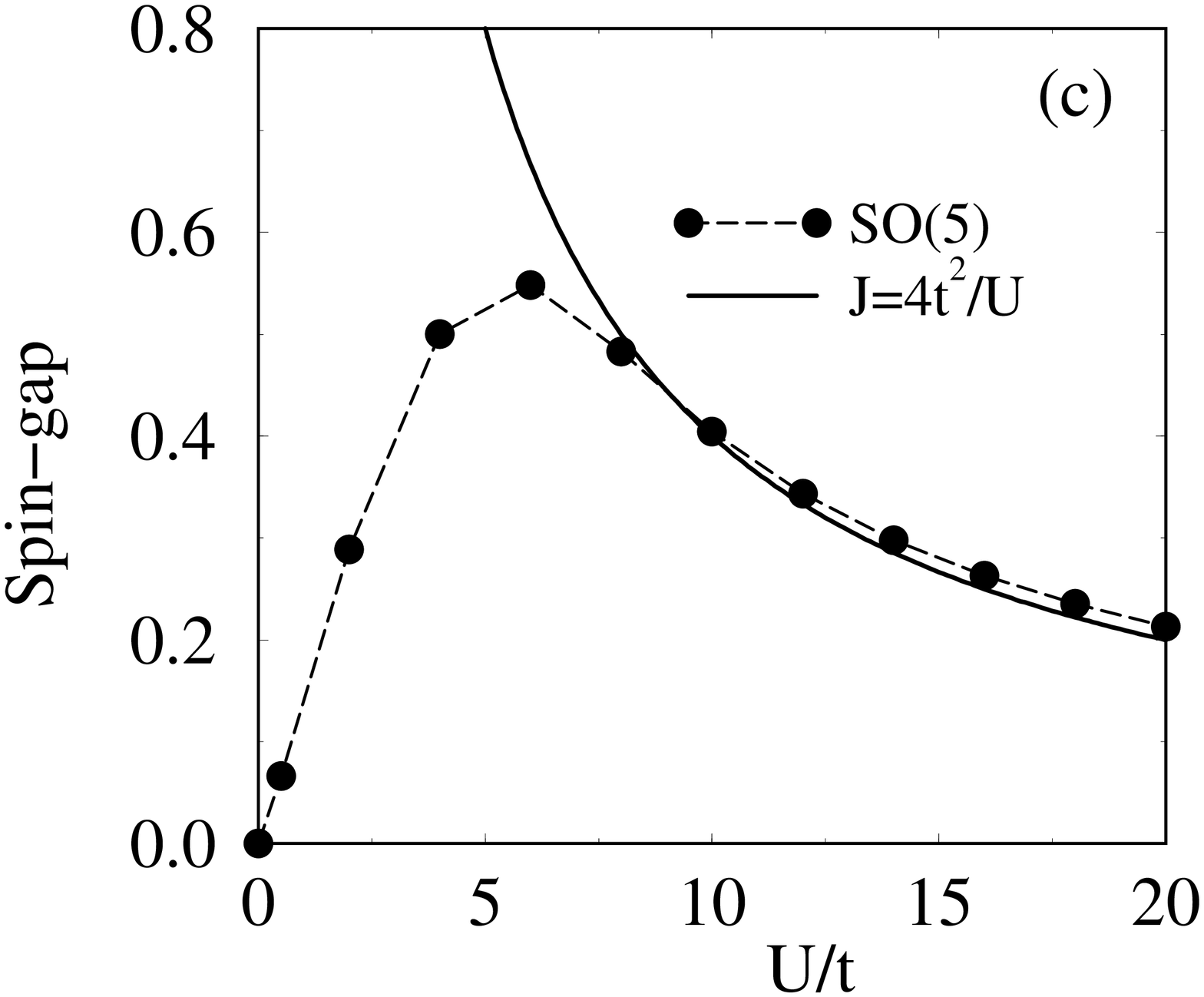,width=7cm,bbllx=98pt,bblly=36pt,bburx=590pt,bbury=700pt,angle=270}}
\vspace{0.25cm}
\caption{(a) Exact diagonalization results for the dynamical spin
response for a $2\times 6$ ladder at half-filling with $U=4t$ as $V$ 
goes from zero up to the $SO(5)$ value of $-4t$. (b) Same as (a) for
$U=8t$ and $V$ varying from zero to $-8t$.
(c) The spin-gap as a function of $U$ for the $SO(5)$
case with $V=-U$.  The solid line is the strong coupling result,
$J=4t^2/U$.}
\label{fig3}
\end{figure}

Because of the large Hilbert space at half-filling, the system sizes
which can be studied with ED are limited, and we will only present
results for a $2\times 6$ ladder.  On the other hand, an advantage of
ED is that it allows for the direct calculation of the spin-gap, as
well as dynamical spectra\cite{meixner,eder} given by:

\begin{eqnarray}
\hat{O}({\bf q},\omega) =  -\frac{1}{\pi}\mbox{Im}
\langle {\bf k}| 
\hat{O}^\dagger_{-{\bf q}}
\frac{1}{\omega+E_{g.s.}-\hat{H}-i\epsilon}
\hat{O}_{{\bf q}}
|{\bf k}\rangle
\end{eqnarray}

\noindent where $|{\bf k}\rangle$ is the ground state wave function
with momentum ${\bf k}$, and $\hat{O}({\bf q},\omega)$ 
is an arbitrary translationally
invariant operator.

What is the nature of the ground state of the $2\times 6$ ladder?  In
Fig. 3(a), the dynamical spin response, using $\hat{O}_{\bf q} =
\hat{S}^z_{\bf q} = \sum_j \mbox{exp}(-i{\bf q\cdot r_j})
\hat{S}^z({\bf r_j})$, is shown for $U=4t$\cite{spread}. Starting from
the Hubbard ladder, $V=0$, we find that a spin-gap is present at
half-filling\cite{noack}.  In particular, as seen in Fig. 3, as $V$
approaches $-U$, the spin-gap remains robust and even increases for
the $SO(5)$ ladder rather than going to zero as one might expect from
the strong coupling ($J\rightarrow 0$) limit.  The low lying spin
excitations are dominated by a large peak in the spin response
occurring at a momentum transfer of $(\pi,\pi)$.  (Note that
Figs. 3(a) and (b) show the momentum integrated spin excitation
spectrum, $S(\omega) = \int d{\bf q} S({\bf q},\omega)$.)
Furthermore, there is little qualitative difference between the
spectra as the model moves toward the ladder of higher symmetry.
Similar behavior is seen for $U=8t$ (Fig. 3(b)) where again a spin-gap
is always present.  An angular resolved examination of $S({\bf
q},\omega)$ shows that the dominant low energy peak in $S(\omega)$ is
due to inter-band scattering processes, involving a momentum transfer
of $\pi$ in the rung direction, whereas the quasi-continuous spectral
weight at higher frequencies stems from intra-band processes.

The spin-gap can also be obtained by measuring the energy difference
between the ground state in the $S^z=0$ sector and the ground state
with total $S^z=1$ at a momentum of $(\pi,\pi)$, shown in Fig. 3(c)
with $V=-U$. The value obtained in this way agrees exactly with the
position of the lowest peak in $S(\omega)$.  The gap increases,
reaching its maximum around the intermediate coupling strength of
$U=6t$, and then falls off as $1/U$.  For comparison, the expected
result from strong coupling given by $4t^2/U$ is shown in Fig. 3(c) as
the solid line.  Therefore, it is understandable why the power law
behavior in the correlations was not seen from the QMC simulations.
An effective spin-spin interaction, $J_{eff}=4t^2/U$, causes the
system to have a spin gap with the ground state being well described
by a product of rung singlets.  Hence, no degeneracy between the two
$SO(5)$ multiplets exists.

In order to explore the approximate symmetry of the Hubbard ladder
and how it evolves as a function of $V$, the dynamic response of the
$\pi$ operator, i.e., $\hat{O}_{\bf q} = \hat{\pi}_{\bf q} = \sum_{\bf r}
\mbox{exp}(-i{\bf q\cdot r}) \hat{\pi}({\bf r})$ with 

\begin{eqnarray}
\hat{\pi}({\bf r}) = \frac{1}{2}\bigl(\hat{\pi}_x({\bf r}) + 
i\hat{\pi}_y({\bf r})\bigr) 
= \frac{-i}{2}(-1)^{r} c_\uparrow({\bf r}+\hat{y})c_\uparrow({\bf r}),
\end{eqnarray}

\noindent was measured for the state with one electron pair more than
the half-filled state.  Fig. 4 shows a direct comparison at $U=4t$ and
$U=8t$ between the low energy spin excitations and the $\pi$ resonance
using the energy of the half-filled ground state as a reference.
Clearly, both have a dominant low energy peak at the same excitation
energy independent of the rung coupling $V$, indicating that the
resulting final states are equivalent.  Thus, an approximate $SO(5)$
symmetry is revealed.  However, it should be noted that the $\pi$
resonance has significant spectral weight at higher energies (not
shown in Fig. 4) for the Hubbard ladder, whereas all the weight shifts
to the low energy peak as the symmetric ladder is
approached\cite{pinote}.  A thorough finite-size scaling analysis is
expected to show that this low energy $\pi$ resonance survives in the
thermodynamic limit as the length of the ladder is increased.

\begin{figure}[tbh]
\vspace{0.25cm}
\centerline{\psfig{figure=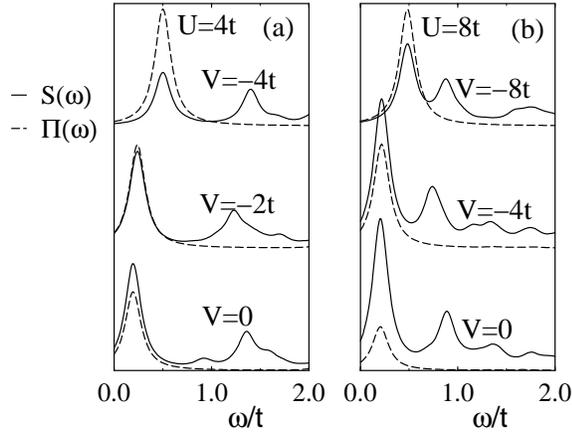,width=7cm,bbllx=98pt,bblly=36pt,bburx=590pt,bbury=700pt,angle=270}}
\vspace{0.25cm}
\caption{(a) Comparison between the low lying spin excitations of a
$2\times 6$ ladder at half-filling with $U=4t$ and the $\pi$ excitation
spectrum of the ground state with one pair of electrons above half-filling.
(b) Same as (a) for $U=8t$.
Note that the energies are taken with respect to the ground state
energy at half-filling.}
\label{fig4}
\end{figure}

In this paper, we have numerically analyzed a model which can be
varied by changing a rung interaction so that it interpolates between
the Hubbard and the $SO(5)$ SZH model on a two-leg ladder.  We find
that the spin-gap vanishes as $U/t$ goes to infinity, and the ground
state is well described by rung singlets for intermediate to large
coupling strengths.  This can be understood since the hopping
along the chain and within a rung creates an effective spin-spin
interaction $J_{eff}=4t^2/U$.  This effective spin-spin
interaction must be overcome by an even stronger negative value of the
rung interaction, $-|V|>U$ in order to restore the degeneracy between
rung singlets and triplets obtained as $U/t\rightarrow\infty$.
Furthermore, no qualitative changes occur in the low lying excitations
of the Hubbard ladder as the $SO(5)$ symmetric point is approached.
Therefore, we conclude that the $\pi$ operators are approximate
eigenoperators of the Hubbard ladder near half-filling, and hence, the
low energy behavior of the Hubbard ladder in the intermediate to
strong coupling range has an approximate $SO(5)$ symmetry.

The authors are deeply grateful to D. Scalapino, and have benefited
from enlightening conversations with A. Sandvik, A. Moreo, R. Noack,
S. R. White, E. Jeckelmann, E. Demler, B. Sugar, C. Martin, A. DeLia,
R. Konik and S.-C. Zhang. D. Duffy acknowledges the support from the
Dept. of Energy under grant DE-FG03-85ER451907.  We thank the San
Diego Supercomputing Center for providing us access to their
facilities.

\vfil


\begin{references}
\bibitem{qmcnote} One reason for imposing the condition $J=0$ is to
simplify the QMC simulation.  The Hubbard-Stratonovich transformation of
the spin-spin interaction results in a large number of auxiliary fields
over which one must sample to obtain the ground state properties.  As
can be seen with just the $V$-term which introduces 4 additional fields,
the sign problem becomes worse when $J$ or $V$ become comparable to
$U$, even at half-filling.
\bibitem{temp} For the intermediate value of the coupling $U=4t$
presented here, we could reach lower temperatures of $\beta t = 3$ or
4.  However, the sign problem grows with decreasing temperature.
\bibitem{spread} The resulting delta function peaks in the spin
response are replaced by Lorentzians of the form
$\delta(\omega)=\frac{1}{\pi}\frac{\epsilon}{\omega^2+\epsilon^2}$.
The broadening used throughout this work was $\epsilon=0.1t$.
\bibitem{pinote} 
We have checked explicitly that at the $SO(5)$ point all of
the weight of the $\pi$ operator goes into the lowest energy peak
shown in Fig. 4.
\end{references}
\end{document}